\documentclass[runningheads]{llncs}
\usepackage{graphicx}
\usepackage{color}
\usepackage{cite}
\usepackage{booktabs}
\usepackage{multirow}
\usepackage{subcaption}
\usepackage{xspace}
\usepackage[misc,geometry]{ifsym}

\definecolor{dkgreen}{rgb}{0,0.6,0}
\definecolor{dimgray}{rgb}{0.41, 0.41, 0.41}
\definecolor{brickred}{rgb}{0.8, 0.25, 0.33}
\definecolor{cadmiumgreen}{rgb}{0.0, 0.42, 0.24}
\definecolor{gray}{rgb}{0.4,0.4,0.4}
\definecolor{javared}{rgb}{0.6,0,0} %
\definecolor{javagreen}{rgb}{0.75,0.5,0.35} %
\definecolor{javapurple}{rgb}{0.5,0.5,0.35} %
\definecolor{javadocblue}{rgb}{0.55,0.35,0.75} %
 
\usepackage{listings}
\lstdefinelanguage{diff}{
  morecomment=[f][\color{dimgray}]{@@},
  morecomment=[f][\color{cadmiumgreen}]{+\ },
  morecomment=[f][\color{brickred}]{-\ },
}

\def\repo{\url{https://github.com/coinse/Defects4J-multifault}}
\def\dfj{Defects4J\xspace}

\begin{document}
\title{Searching for Multi-Fault Programs in Defects4J}
\author{Gabin An\orcidID{0000-0002-6521-8858}\and
Juyeon Yoon\orcidID{0000-0003-2706-1156}\and
Shin Yoo\textsuperscript{(\Letter)}\orcidID{0000-0002-0836-6993}}
\authorrunning{G. An et al.}
\institute{KAIST, Daejeon, Republic of Korea\\
\email{\{agb94,juyeon.yoon,shin.yoo\}@kaist.ac.kr}}

\maketitle              %
\begin{abstract}
\dfj has enabled numerous software testing and debugging research work since 
its introduction. A large part of its contribution, and the resulting 
popularity, lies in the clear separation and distillation of the root cause of 
each individual test failure based on careful manual analysis, which in turn 
allowed researchers to easily study individual faults in isolation. However, in 
a realistic debugging scenario, multiple faults can coexist and affect test 
results collectively.
Study of automated debugging techniques for these situations, such as
failure clustering or fault localisation for multiple faults, would 
significantly benefit from a reliable benchmark of multiple, coexisting faults. 
We search for versions of \dfj subjects that contain multiple faults, by 
iteratively transplanting fault-revealing test cases across \dfj versions. 
Out of 326 studied versions of \dfj subjects, we report that over 95\% (311 
versions) actually contain from two to 24 faults. We hope 
that the extended, multi-fault \dfj can provide a platform for future research 
of testing and debugging techniques for multi-fault programs.

\end{abstract}

\section{Introduction}
\label{sec:introduction}
\dfj~\cite{just2014defects4j} is one of the most popular real-world Java fault datasets in the field
of software engineering, with over 650 citations as of June 2021 since its publication in 2014.
\dfj provides a number of software faults, along with a clearly separated and isolated set of test
cases that can reveal each fault, making it easier for researchers to study individual faults in
isolation. Due to both the ease of use and the realism of the curated faults, it has been broadly adopted in the empirical validation of numerous automated
debugging work such as Fault Localisation (FL)~\cite{b2016learning, sohn2017fluccs, li2019deepfl} and Automated Program Repair (APR)~\cite{chen2019sequencer, liu2019tbar, koyuncu2020fixminer}.

However, in realistic debugging scenarios, multiple faults can coexist in software and affect the test
results together. For example, a Continuous Integration (CI) process of large-scale industry software can produce hundreds of failing test cases
that are caused by distinct root causes~\cite{golagha2019failure}.
The isolation of individual faults that made \dfj compatible with the Single Fault Assumption (SFA) ironically prevents it from being used to study the debugging of multiple faults. 
According to a systematic literature review of multiple faults localisation~\cite{zakari2020multiple},
the majority (33) of the 55 selected studies used only C faults
for the evaluation. Only ten studies are reported to consider Java programs, five out of which employ \dfj~\cite{laghari2016fine,zheng2018localizing,xia2016automated,zhang2017boosting,li2016iterative}. Only Zheng et al.~\cite{zheng2018localizing}
combined separate multiple \dfj faults; since the procedure of creating the multiple faults was manual, only 46 have been created. The remaining work either concern multi-hunk faults, i.e., a single fault that can only be fixed by changing multiple locations~\cite{saha2019harnessing} and consequently use \dfj as it is~\cite{zhang2017boosting,xia2016automated}, or actually concern neither multiple faults nor multi-hunk faults~\cite{laghari2016fine,li2016iterative}. Note that, in this paper, we use the term \emph{multiple faults} to denote the faults that can be fixed independently of each other.

Given the contributions to the automated debugging research made by \dfj under SFA, we believe that the study of automated 
multi-fault debugging techniques~\cite{zakari2020multiple},
such as failure clustering~\cite{jones2007debugging,golagha2019failure,dang2012rebucket} or fault localisation for multiple faults~\cite{
abreu2009spectrum, zheng2018localizing, ghosh2021spectrum}, would significantly benefit from the construction of a 
reliable dataset of realistic multi-fault Java programs. In this paper, we 
build a \textbf{real-world} Java \textbf{multi-fault} dataset by extending 
\dfj. Instead of artificially injecting mutation or manually grafting faults, 
we use iterative search to systematically detect the existence of multiple faults in each version via fully automated transplantation and execution of the fault-revealing test cases. We report that 311 out of 326 studied faulty versions (95.4\%) contain multiple faults, ranging from two to 24. The result data and replication package are publicly available\footnote{\repo}.

\section{Proposed Approach\label{sec:approach}}

The faults in \dfj are extracted from the actual development history of various projects. Since every fault has a different life span~\cite{canfora2011long,kim2006long},
even a fault that was recently fixed may have existed in the project for a long time. In this work, we check if a specific fault $N$ in version $P$ of a \dfj subject exists in an older version $P'$ containing another fault $M$. If $N$ exists in $P'$, we regard $P'$ as a multi fault program that includes both $N$ and $M$. Note that we modify neither $P$ nor $P'$: the check is performed by test transplantation, and therefore we only reveal what already exists in $P'$.
The following sections present the motivating example and our proposed method to search for multi-fault programs.

\subsection{A Motivating Example}

\begin{lstlisting}[language=diff,float=t,caption={The developer patch for Math-5},
  label={lst:Math-5b},captionpos=b]
--- a/src/main/java/org/apache/commons/math3/complex/Complex.java
+++ b/src/main/java/org/apache/commons/math3/complex/Complex.java
@@ -304,7 +304,7 @@ @@ public Complex reciprocal() {
        if (real == 0.0 && imaginary == 0.0) {
-           return NaN;
+           return INF;
        }
\end{lstlisting}

\begin{lstlisting}[language=Java,float=t,caption={The fault-revealing test case of Math-5},
label={lst:testReciprocalZero},captionpos=b]
public void testReciprocalZero() {
    Assert.assertEquals(Complex.ZERO.reciprocal(), Complex.INF);
    // Error message: junit.framework.AssertionFailedError: expected:<(NaN, NaN)> but was:<(Infinity, Infinity)>
}
\end{lstlisting}

\begin{lstlisting}[language=Java,firstnumber=304,float=t,caption={In Math-6b, \texttt{Complex.java} (line 305) contains the fault Math-5},
label={lst:Math-6b},captionpos=b]
        if (real == 0.0 && imaginary == 0.0) {
            return NaN; // Math-5b
        }
\end{lstlisting}

Listing~\ref{lst:Math-5b} shows the fault Math-5 in \dfj and its developer patch
changing the return value from \texttt{NaN} to \texttt{INF}.\footnote{\url{http://program-repair.org/defects4j-dissection/\#!/bug/Math/5}}
This fault is revealed by the test case \texttt{testReciprocalZero}
(Listing~\ref{lst:testReciprocalZero}) that checks if the return value is equal to \texttt{INF}.
Each \dfj fault is similarly provided with a set of fault-revealing test cases that reveals a single fault.

We note that, with few exceptions of recently added subjects and versions, the majority of faulty versions in \dfj are indexed chronologically based on their revision dates, so that a lower ID refers to a more recently fixed fault:  
for instance, Math-5 was fixed later than Math-6. 
Therefore, the faulty source code version of Math-6 (referred to as Math-6b) may also contain the fault Math-5. Listing~\ref{lst:Math-6b} confirms that Math-5 does exist in Math-6b, but is simply not revealed 
due to the absence of the fault-revealing test case, \texttt{testReciprocalZero}.
When transplanted to Math-6b, the test fails with the same error message as in Math-5b, showing that Math-6b contains at least two faults, Math-5 and Math-6.

\subsection{Searching For Multiple Fault Versions}\label{sec:search}

Let $B_{M}$ be the \dfj faulty source code version that corresponds to the fault $M$.\footnote{\texttt{defects4j checkout -p Math -v 6b -w <dir>} checks out $B_{Math-6}$ into \texttt{<dir>}.}
As shown in our motivating example, if a fault $N$ is fixed after a fault $M$,
the fault $N$ may \textit{already exist} in $B_M$. Consequently, to build a multi-fault dataset, we check which faults exist in which preceding faulty versions.

\paragraph{\textbf{Search Strategy}}
For each fault $N$ in a project, we sequentially check whether the fault exists in each previous faulty version $B_M$, such that $M.id > N.id$, from the latest version to the older version.
The search stops once $N$ is not revealed in $B_M$. For example, the fault Lang-3 is revealed in
Lang-[4,16]b, but not in Lang-17b. In this case, the search immediately stops and moves to the next iteration with a new $N$ (Lang-4).
This is because if $B_M$ does not contain the fault $N$, it is likely that versions older than $B_M$ do not include $N$ either.

\paragraph{\textbf{Existence Check}}
To determine the presence of a fault N in $B_M$,
we \textit{transplant} all fault-revealing tests of $N$ to
$B_M$. We confirm that $N$ exists in $B_M$ if and only if (1) all
target test class files to where test case methods are transplanted exist in $B_M$, (2) all transplanted test cases are successfully compiled and fail against 
$B_M$, and (3) the error messages in $B_M$ are the same as those in $B_N$.
If the fault-revealing test cases of the faults $N$ and $M$ overlap with each other, 
we further execute the fault-revealing tests of $N$ on the fixed version of $M$ to 
ensure that the overlapped test cases still fail due to $N$ without the 
presence of $M$.

\paragraph{\textbf{Building Multi-fault Subjects}}
When the above search is done, we obtain the set of pairs $E$ such that
$(N, M) \in E$ if and only if $N$ exists in $B_M$. %
For every fault $M$ in \dfj, the set of \textit{found faults} in $B_M$, $F(B_M)$, is defined as
${F(B_{M}) = \{M\} \cup \{N|(N, M) \in E\}}$.
If $|F(B_{M})| > 1$, $B_M$ is a multi-fault subject.

\subsection{Implementation Details}
The process in Section~\ref{sec:search} is dockerised and automated.
We use \texttt{javaparser}~\footnote{\url{https://github.com/javaparser/javaparser}} to detect the line range of the target test methods during transplantation. In the docker container, one can simply checkout to the multi-fault version by invoking
\texttt{python3.6 checkout.py Math-1-2-3 -w /tmp/Math-1-2-3}, 
after which the 
\textit{same} source code with Math-3b, augmented with the fault-revealing test cases of Math-1 and Math-2, is checked out.

\begin{figure}[t]
  \begin{subfigure}[t]{0.48\textwidth}
      \centering
      \includegraphics[width=\textwidth]{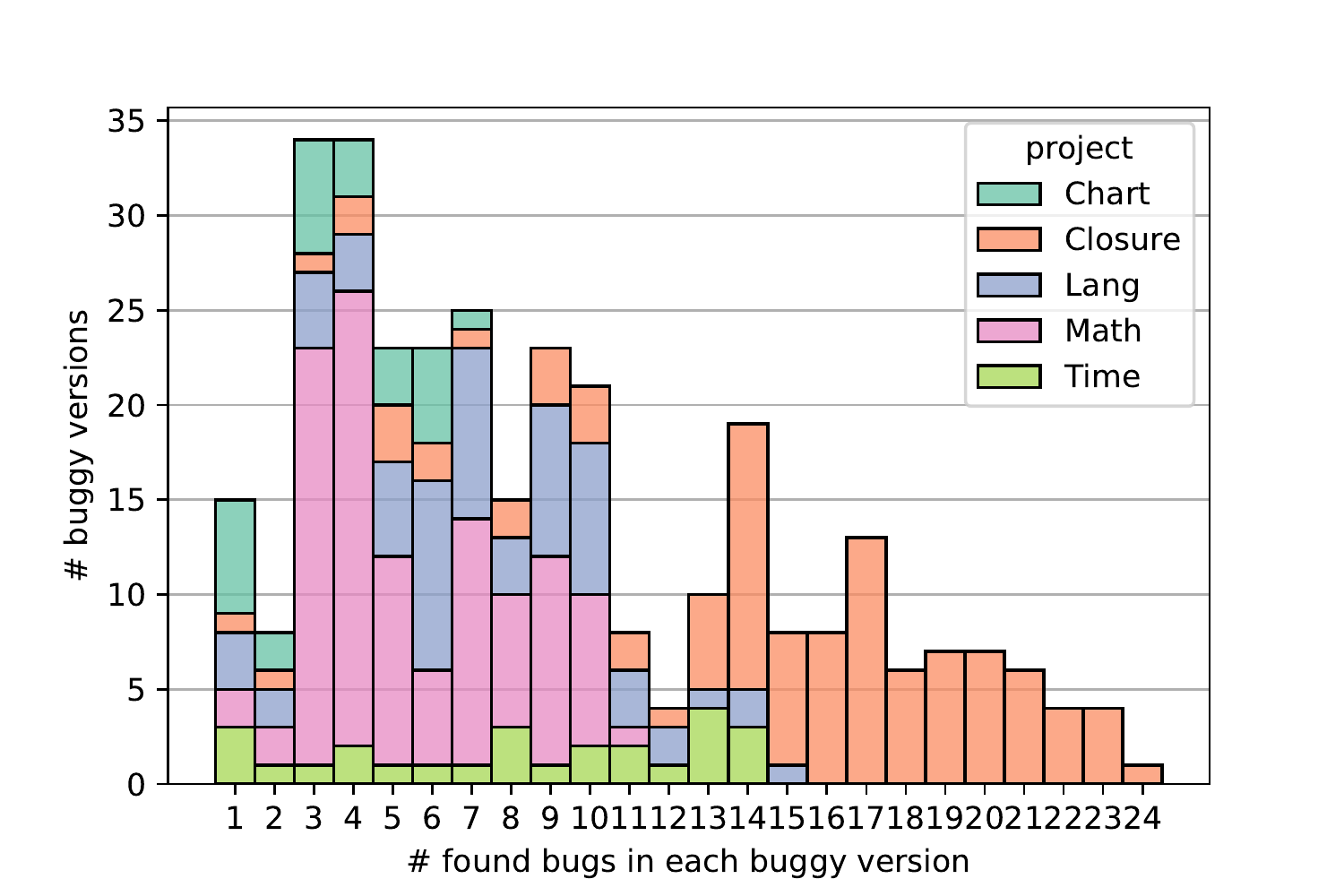}
      \caption{The number of faulty versions in \dfj with each number of faults}
      \label{fig:stats}
  \end{subfigure}
  \hspace{10pt}
  \begin{subfigure}[t]{0.48\textwidth}
      \centering
      \includegraphics[width=\textwidth]{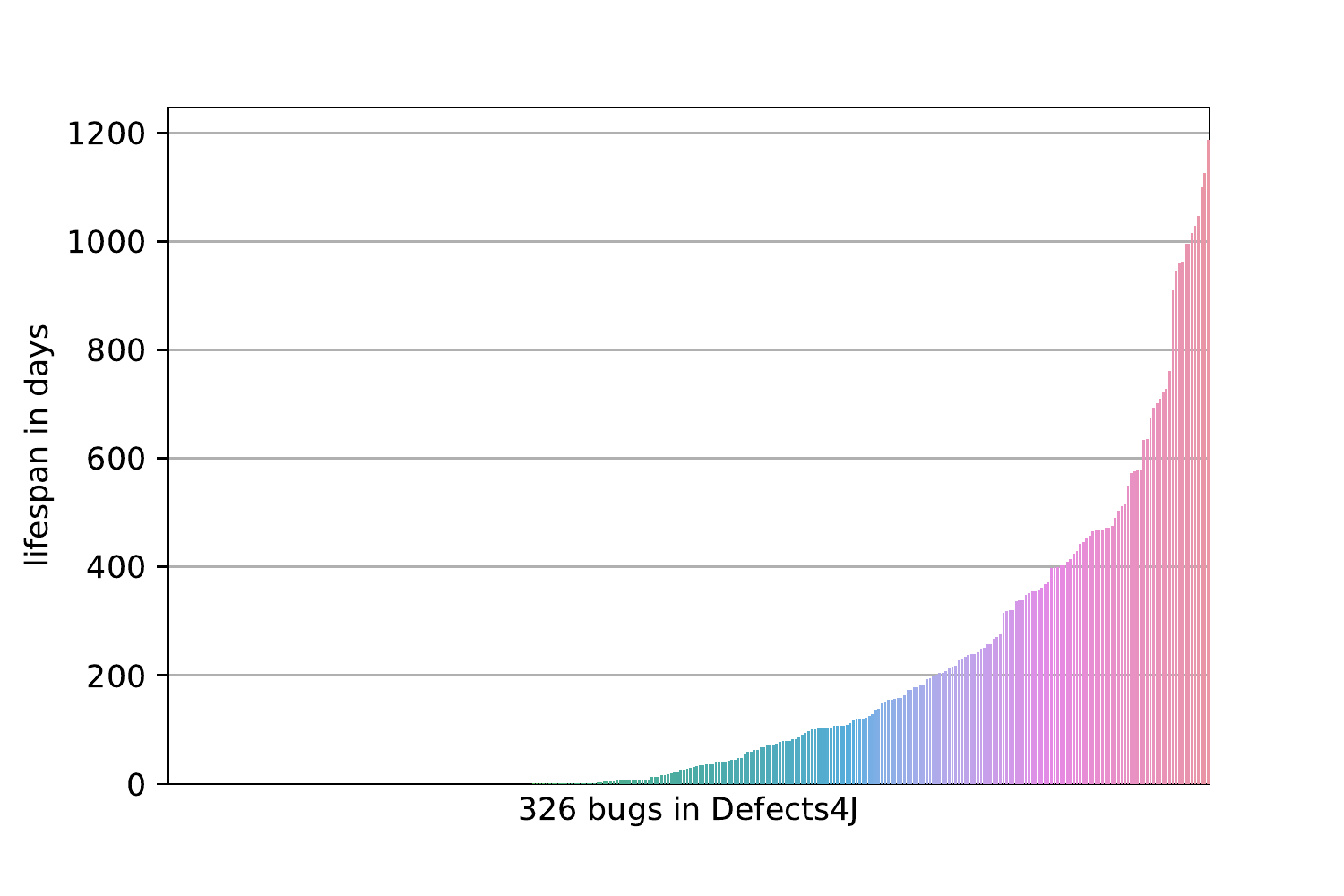}
      \caption{The sorted life span of faults in days (average=154, standard deviation=246)}
      \label{fig:lifespan}
  \end{subfigure}
  \caption{The summary of search results}
\end{figure}

\section{Results}

\paragraph{\textbf{Multiple Fault Subjects}}
Figure~\ref{fig:stats} shows how many faults are contained in the faulty versions of five projects\footnote{\dfj Bug IDs: Lang 1-65, Chart 1-26, Math 1-106, Time 1-27, and Closure 1-106.
Note that Lang-2, Time-21, Closure-63 and -93 are excluded since they are either no longer reproducible under Java 8 or the duplicate bugs.}.
The x-axis shows the number of faults found in each faulty version, and the y-axis shows the number of faulty versions.
Out of 326 faulty programs, 95.4\% (=311/326) of them contain multiple faults (i.e., \# found faults $>$ 1).
Furthermore, 126 and 22 faulty versions have $\geq 10$ and $\geq 20$ faults, respectively. For example, Closure-90b contains 24 faults. Our repository contains the full results of the found multi-fault versions.

\paragraph{\textbf{Lifespan of Faults}}
To confirm whether lifespans of \dfj faults vary similary to existing 
findings~\cite{kim2006long,canfora2011long}, we calculate the lifespan of each fault.
Let us define the \textit{lifespan} of fault $N$ as the number of days between the date of the oldest previous faulty version where fault $N$ is detected and the revision date of $N$ when the patch is applied.
If there is no preceding version where the fault $N$ is revealed, the lifespan is zero.
Figure~\ref{fig:lifespan} shows that lifespans of faults range from 0 days up to longer than three years (e.g., Lang-41 has the lifespan of 1,187 days). The  variance in lifespan suggests that the probability of having multiple faults at any given time can be nontrivial.

\section{Conclusion}
The paper presents a multi-fault Java dataset based on \dfj, for which subjects with multiple real faults are constructed by transplanting tests without modifying the source code. Exploiting the chronological indexing of \dfj, we propose a systematic search strategy to find co-existing faults that have not yet been revealed by failing tests. The results show that 311 out of 326 versions in \dfj actually contain multiple faults.  
We hope that our extension of \dfj can aid future research on search-based automated debugging under the existence of multiple faults. 

\section*{Acknowledgement}
This work is supported by National Research Foundation of Korea 
(NRF) Grant (NRF-2020R1A2C1013629), Institute for Information \& communications Technology Promotion grant funded by the Korean government (MSIT) (No.2021-0-01001), and Samsung Electronics (Grant No. IO201210-07969-01).

\bibliographystyle{splncs04}
\bibliography{ref}

\end{document}